\documentclass[12pt,preprint]{aastex}
\usepackage{hyperref}
\usepackage{amsmath}
\usepackage{color}
\usepackage{bm}

\begin{document}

\title{An Outer Arm in the Second Galactic Quadrant: Structure}

\author{Xinyu~Du\altaffilmark{1,2,3}, Ye~Xu\altaffilmark{1,3}, Ji~Yang\altaffilmark{1,3}, Yan~Sun\altaffilmark{1,2,3},
Facheng~Li\altaffilmark{1,2,3}, Shaobo~Zhang\altaffilmark{1,3}, Xin~Zhou\altaffilmark{1,3}}
\email{xydu@pmo.ac.cn; xuye@pmo.ac.cn}

\altaffiltext{1}{Purple Mountain Observatory, Chinese Academy of Science, Nanjing 210008, China}
\altaffiltext{2}{Graduate University of the Chinese Academy of Sciences, 19A Yuquan Road, Shijingshan District,
Beijing 100049, China}
\altaffiltext{3}{Key Laboratory of Radio Astronomy, Chinese Academy of Science, Nanjing 210008, China}

\begin{abstract}
The lack of arm tracers, especially the remote tracers,
is one of the most difficult problems preventing us from studying the structure of the Milky Way.
Fortunately, with its high-sensitivity CO survey,
the Milky Way Imaging Scroll Painting (MWISP) project offers such an opportunity.
Since completing about one third of its mission,
an area of $l=[100,150]^{\circ}$, $b=[-3,5]^{\circ}$ has nearly been covered.
The Outer arm of the Milky Way first clearly revealed its shape
in the second galactic quadrant in the form of molecular gas ---
this is the first time that the Outer arm has been reported in such a
large-scale mapping of molecular gas.
Using the 115 GHz $^{12}$CO(1-0) data of MWISP at the LSR velocity $\simeq[-100,-60]$ km s$^{-1}$
 and in the area mentioned above,
we have detected 481 molecular clouds in total,
and among them 332 (about 69\%) are newly detected
and 457 probably belong to the Outer arm.
The total mass of the detected Outer arm clouds is $\rm \sim 3.1\times10^{6} \ M_{\odot}$.
Assuming that the spiral arm is a logarithmic spiral, the pitch angle is fitted as $\rm \sim13.1^{\circ}$.
Besides combining both the CO data from MWISP and the 21 cm HI data from the Canadian Galactic Plane Survey (CGPS),
the gas distribution, warp, and thickness of the Outer arm are also studied.
\end{abstract}


\section{Introduction}\label{Sec: Introduction}
It is well-known that our galaxy has a spiral structure,
\citep{1958MNRAS.118..379O,1959Obs....79...58B}
but a detailed understanding of the Milky Way is still lacking.
Since \citet{1976A&A....49...57G} presented a large-scale spiral pattern with HII regions,
nearly 100 Milky Way models have been proposed \citep{2010gama.conf...45S}.
However, most models share the same large structure in spite of differing details 
(e.g., \citealt{2007A&A...470..161R}; \citealt{2008AJ....135.1301V}; \citealt{2014A&A...569A.125H}).

The fact that there exists a spiral arm beyond the Perseus arm 
has been generally recognized for a long time. 
Since \citet*{1975A&AS...20...85M} noticed that several young star clusters 
were located far beyond the position of the Perseus arm, 
more work has been done to confirm the existence of this external arm. 
\citet{1982ApJ...263..116H} first systematically studied the distribution 
of HI in the outer galaxy beyond the solar circle, 
which is a vast frontier of our galaxy. 
\citet{1998ApJS..115..241H} described the molecular image of outer galaxy 
with FCRAO CO survey results, 
providing the evidence for the presence of molecular clouds in this outer area. 
\citet{2003A&A...406..119N} traced the external arm using OB stars
and concluded that the Cam OB3 association lies on it.
On the basis of maser source (associated with high-mass star-forming region, HMSFR) distances, 
\citet{2014ApJ...783..130R} first delineated several spiral arms, 
including the Outer arm, using the trigonometric parallax method.
In addition, detailed parallax results and analyses of the Outer arm are also presented by \citet{2015ApJ...800....2H}

However, the accurate maser source positions cannot outline the distributions of interstellar gases.
The FCRAO CO survey only traced the high-mass molecular clouds because of its poor sensitivity.
Fortunately, the Milky Way Imaging Scroll Painting (MWISP) project
\footnote{http://english.dlh.pmo.cas.cn/ic/ or http://www.radioast.nsdc.cn/mwisp.php}
provides such a chance to study the molecular gases
with the no-bias high sensitive $^{12}$CO($1-0$), $^{13}$CO($1-0$) and C$^{18}$O($1-0$) observations.
Due to such high-sensitive observations,
a new spiral arm (hereafter the New arm) beyond the Outer arm has been discovered by \citet{2015ApJ...798L..27S}.
Combining the $^{12}$CO($1-0$) observations of MWISP and
atomic hydrogen data from the Canadian Galactic Plane Survey (CGPS; \citealt{2003AJ....125.3145T}),
we presented the results of the Outer arm:
the arm located between the Perseus arm and the New arm.
The results derived from $^{13}$CO($1-0$) and C$^{18}$O($1-0$) data and further analyses
will be published in our future works.

Please note that the Outer arm does not have a particular name yet.
Some authors have labeled it as the ``Cygnus arm'', or the ``Perseus +I arm'', 
or the ``Norma---Cygnus arm'' \citep{2008AJ....135.1301V}. 
Otherwise, the name ``Outer arm'' is also widely used \citep[e.g.,][]{2001ApJ...547..792D,2014ApJ...783..130R}. 
Considering the fact that the name ``Cygnus'' has been described as 
the location near the Sun in some early papers,
here we choose the name ``Outer arm'' in this paper to avoid confusion.

In Sect. \ref{Sec: Observation and archival data}
we introduce our CO observation conditions and archival data of atomic hydrogen. 
In Sect. \ref{Sec: Analysis of clouds} we describe how we pick out the Outer arm clouds
and briefly study their properties.
In Sect. \ref{Sec: Properties of Outer arm} we study the properties of the Outer arm,
including the pitch angle, the gas distribution, the thickness, and the warp. 
The summary is given in Sect. \ref{Sec: Summary}.

\section{Observations and archival data}\label{Sec: Observation and archival data}
\subsection{CO observations}\label{SubSec: CO observation}
The $^{12}$CO ($1-0$), $^{13}$CO($1-0$) and C$^{18}$O($1-0$) lines 
were observed simultaneously using the Purple Mountain Observatory 
Delingha (PMODLH) 13.7 m telescope from 2011 September to 2015 March
as one of the scientific demonstration regions for the MWISP project,
which is the first no-bias high-sensitivity CO survey
with such a large-scale aiming at $l=[-10,250]^{\circ}$, $b=[-5,5]^{\circ}$.
With the on-the-fly (OTF) observing mode, 
MWISP now has completed about one-third of its plan,
and an area of $l=[100,150]^{\circ}$, $b=[-3,5]^{\circ}$ has mostly been covered
---the total area was 288 square degrees. 
A superconductor-insulator-superconductor (SIS) superconducting receiver with a nine-beam array
was used as the front end \citep{6313968}. 
A Fast Fourier Transform (FFT) spectrometer with a total bandwidth of 
1000 MHz and 16,384 channels was used as the back end.
For the 115 GHz $^{12}$CO ($1-0$) observations,
the main beam width was about $52''$, the main beam efficiency ($\eta_\mathrm{MB}$) was 0.46,
and the typical rms noise level was $\sim$ 0.5 K, corresponding to a channel width of 0.16 km s$^{-1}$.
($\sim$ 0.2 K per 0.8 km s$^{-1}$ channel, which is 3 times better than the FCRAO OGS sensitivity.)
All the data were corrected by $T_\mathrm{MB}=T^{*}_\mathrm{A}/\eta_\mathrm{MB}$.
The data were sampled every $30''$.
All the data were reduced using the GILDAS/CLASS package.

\subsection{Archival data of atomic hydrogen}\label{SubSec: Archival data of atomic hydrogen}
The 21 cm line data were retrieved from the CGPS.
We downloaded data of $l=[63,155]^{\circ}$, $b=[-3,5]^{\circ}$
from the Canadian Astronomy Data Centre\footnote{http://cadc.hia.nrc.ca}.
The velocity coverage of the data is in the range of -153 to 40 km s$^{-1}$, with a channel separation of 0.82 km s$^{-1}$.
The survey has a spatial resolution of $58''$, which is comparable to our CO observations.

\section{Analysis of clouds}\label{Sec: Analysis of clouds}
We have detected 481 clouds in total;
332 (about 69\%) clouds are newly detected,
457 clouds are identified in the Outer arm,
and 24 are indentified in the New arm.
Among the Outer arm clouds, 7 are reported by \citet{1994A&AS..103..503B} (hereafter BW94),
75 are reported by \citet{2001ApJ...551..852H} (hereafter HCS01 clouds),
125 are reported by \citet{2003ApJS..144...47B} (hereafter BKP03).
And all 24 of the New arm clouds are newly detected.
(These 24 New arm clouds do not overlap with the clouds detected by \citealt{2015ApJ...798L..27S}.)
The parameters of all the 481 clouds are summarized in Table.\ref{Table}.

\subsection{Cloud identification}\label{SubSec: Cloud identification}
Because of the differential rotation of the Milky Way,
most LSR velocities that are consistent with circular Galactic rotation are negative in the second quadrant.
Starting from $V_\mathrm{LSR} \sim 0$ km s$^{-1}$, increasingly negative velocities successively trace
the Local arm, the Perseus arm, the Outer arm, and the New arm.
In order to find all of the Outer arm clouds, we need to know the Outer arm LSR velocity range at every galactic longitude.
As mentioned above, the Outer arm LSR velocity is located between the Perseus arm and the New arm,
which provides us with a clue for picking out the clouds.
First, using the HI data from the CGPS, we plotted the longitude-velocity map of HI,
integrated over all latitudes (from $-3^{\circ}$ to $5^{\circ}$).
Second, we projected the spatial ($x,y$) curves of the Outer arm, the Perseus arm
(both fitted by \citealt{2014ApJ...783..130R},
and hereafter the Reid Outer spiral and the Reid Perseus spiral),
and the New arm (fitted by \citealt{2015ApJ...798L..27S}, and hereafter the Sun New spiral)
on that HI longitude-velocity map.
(In other words, we converted the dashed curves shown in Fig. \ref{Fig: xymap-ab}
into the ones shown in Fig. \ref{Fig: l-v map}.
And the converting method is presented in detail in Sect. \ref{SubSec: l-v, v-b, l-b map}.)
And then we marked all the HCS01 clouds on that map.
Combining the HI map, the positions of the longitude-velocity curves, and HCS01 clouds, 
we can estimate the velocity range of the Outer arm as a function of galactic longitude.
Third, we compiled an automatic procedure to 
list all the positions with emissions greater than 2.5$\sigma$ of the data cube at that velocity range
(where $\sigma$ is the typical rms noise and = 0.5 K).
Last, we checked both the list and data cube and identified the clouds via the naked eye.
It is necessary to emphasize that we did not separate the isolated clouds or cloud complexes
into small pieces of molecular clumps.
In other words, some of the clouds have multiple spectral or spatial peaks.

Finally, we have detected 481 clouds in total.
But not all of them belong to the Outer arm.
Of these clouds, 24 clouds may be located in the New arm,
since the velocity gap between the 24 clouds and the other ones is relatively large in the zoomed-in longitude-velocity map.

However, it is necessary to point out that the Outer arm clouds that we found do not absolutely belong to that arm.
From $l\simeq100^{\circ}$ to $120^{\circ}$, there exists an arm-blending region.
The LSR velocities of HI gas of the Perseus arm and the Outer arm are mixed together
(see Fig. \ref{Fig: l-v map} and \ref{Fig: b-v map}).
Also, the LSR velocity gap of CO between the two arms is not obvious.
Since our cloud identification criterion is based on the LSR velocity,
it is difficult to pick out Outer arm clouds in that region.
We may omit some clouds with an LSR velocity $\gtrsim -70$ km s$^{-1}$,
or falsely pick the clouds that may be located at the inter-arm area.

\subsection{Cloud parameters}\label{SubSec: Cloud parameters}
Heliocentric distance is one of the most important values for studying both the cloud properties and the spiral structure.
Generally there are three methods for measuring the distance:
trigonometric parallax, photometry, and kinematic method.
And the accuracy of the trigonometric parallax and photometry methods
is better than that of the kinematic method.
(see \citealt{2006Sci...311...54X})
However, there is only one source with a parallax distance and a luminosity distance in the Outer arm region:
the one associated with MWISP G135.267+02.800
---its parallax distance is 6.0 kpc \citep{2014ApJ...783..130R}
and its luminosity distance is also 6.0 kpc \citep{2014A&A...569A.125H}.
So the kinematic method is the only choice for us.

We chose to use the kinematic distances
on the basis of the galactic parameters of Model A5 of \citet{2014ApJ...783..130R}
($ \Theta_{0}=240$ km $\rm s^{-1}$, $ R_{0}=8.34$ kpc, $ \frac{d\Theta}{dR}=-0.2$ km $\rm s^{-1}$ $\rm kpc^{-1}$, 
$ U_{\odot}=10.7$ km $\rm s^{-1}$, $ V_{\odot}=15.6$ km $\rm s^{-1}$, $ W_{\odot}=8.9$ km $\rm s^{-1}$, 
$ \overline{U}_{s}=2.9$ km $\rm s^{-1}$, $ \overline{V}_{s}=-1.6$ km $\rm s^{-1}$, 
hereafter Reid model)
and the FORTRAN source code provided by \citet{2009ApJ...700..137R}.
Note that all of the cloud distances we finally used
are kinematic distances, including MWISP G135.267+02.800.

The Reid model is much better than the IAU model
(namely $ \Theta_{0}=220$ km $\rm s^{-1}$,$ R_{0}=8.5$ kpc ).
Take the cloud MWISP G135.267+02.800 as an example.
Its parallax and luminosity distances both are 6.0 kpc,
the Reid kinematic distance (using Reid model) is 6.7 kpc,
whereas the IAU kinematic distance (using IAU model) is 8.6 kpc.
A detailed comparison can be seen in the Section 4 of \citet{2009ApJ...700..137R}

However, one should also keep in mind that there still exist the biases
from the assumption of circular motion inherent in this kinematic distance model.
In the second galactic quadrant the kinematic distance may be a little larger than the real distance,
just as seen in the example of MWISP G135.267+02.800 or the left panel of Fig. \ref{Fig: xymap-ab}.
A detailed discussion about the biases is presented in Sect. \ref{SubSec: Plan view}.

Knowing the distance $d$ and latitude $b$ we calculated the scale height $Z$ by $Z=d\sin(b)$.
And using $b$, $Z$, $d$ and longitude $l$, we calculated the galactocentric radius $R$ with
$R^{2}=d^{2}\cos^{2}(b)+R^{2}_{0}-2R_{0}d\cos(b)\cos(l)+Z^{2}$.

The cloud solid angle $A$ is defined by the 3$\sigma$ limits.
The cloud diameter $D$ is obtained after the beam deconvolution:
$D=d\sqrt{\frac{4}{\pi}A-\theta^{2}_{\mathrm{MB}}}$
\citep{1994ApJ...433..117L},
where $\theta_{\mathrm{MB}}$ is the main beam width.   
Adopting the CO-to-$\rm H_{2}$ $X$ factor
$\rm 1.8 \times 10^{20} cm^{-2} (K \cdot km \ s^{-1})^{-1}$ \citep{2001ApJ...547..792D},
cloud mass is calculated from $ M=2\,\mu\, m_{\mathrm{H}}\, X\, \pi\, (\frac{D}{2})^{2} \int T_{\mathrm{B}} d \,V $ ,
where $ \mu=1.36$ \citep{1983QJRAS..24..267H} is the mean atomic weight per H atom in the ISM,
$ m_{\mathrm{H}}$ is the H atomic mass,
and $\int T_{\mathrm{B}} d \,V$ is the integrated intensity of the $ T_{\mathrm{peak}}$ spectrum.
However, the mass is probably underestimated since the $X$ factor adopted is measured in the solar neighborhood,
and a recent study has suggested an increase of $X$ in the outer Galaxy. \citep{2010ApJ...710..133A}

\subsection{Comparison with HCS01}\label{SubSec: Comparison with hcs01}
\begin{figure}[htbp]
\epsscale{0.8}
\plotone{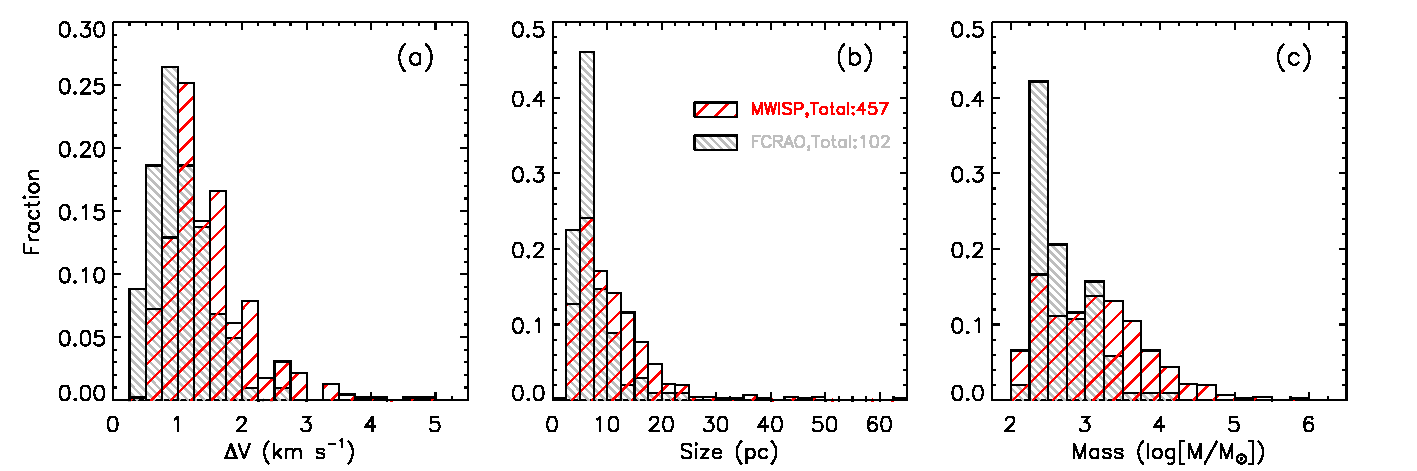}
\caption{Distribution of the Outer arm cloud line width (a), size (b), and mass (c).
The red and gray shades indicate the results from MWISP and FCRAO respectively.
}\label{Fig: MWISP vs FCRAO}
\end{figure}

\citet{2001ApJ...551..852H} have used the CO data of the FCRAO Outer Galaxy Survey
to identify molecular clouds.
To make a comparison, we plotted the distributions of line width, size, and mass for the HCS01 Outer arm clouds and ours.
We have two reasons for not comparing the clouds with BW94 and BKP03:
(i) the number of BW94 clouds is too few 
(ii) the BKP03 clouds are generated by the same data but from a different method than the HCS01 clouds,
and their method tends to find out small-size molecular clumps,
which is not consistent with our cloud identification criterion.
Fig.\ref{Fig: MWISP vs FCRAO} shows the results.
One may notice that there are 102 clouds of HCS01 in Fig. \ref{Fig: MWISP vs FCRAO}, but only 75 clouds are labeled in Table. \ref{Table}.
The reason is the different cloud identification methods:
HCS01 includes both small, isolated clouds and clumps within larger cloud complexes,
but our catalog only includes isolated clouds and cloud complexes.
So the clouds are not exactly matched one-to-one.

Since the cloud heliocentric distances of HCS01 are derived from the IAU model,
the sizes and masses are correspondingly biased.
In order to make our comparison using the same criterion, 
we revised the distances of the Reid model and the sizes and masses are correspondingly revised.

However, the comparison result is against expectations. 
The signal-to-noise ratio of our data is higher than that of HCS01.
Consequently, HCS01 should have contained more luminous ( and therefore more massive ) clouds.
But the distribution shows that the fractions of small sizes and low masses of HCS01 are higher than those of ours.
Two main reasons may have caused this:
(i) HCS01 includes smaller clumps within larger cloud complexes, but we did not detach clumps from cloud complexes.
(ii) Because of our higher signal-to-noise ratio, we can detect a larger angle area for the same cloud, 
which results in larger sizes and masses.

\section{Properties of Outer arm}\label{Sec: Properties of Outer arm}
\subsection{Pitch angle}\label{SubSec: Pitch angle}
Spiral arms of external galaxies are usually approximated by logarithmic spirals.
This is a simple but reasonable assumption and is also applicable to our Milky Way galaxy.
That the pitch angle of one spiral is constant is one of the properties of a logarithmic spiral.
Usually, the spiral arms of one galaxy are roughly fitted by logarithmic spirals with the same pitch
(e.g., \citealt{2008AJ....135.1301V}),
or more accurately, different spiral arms are fitted by different pitches
(e.g., \citealt{2003A&A...397..133R}).
And sometimes even one spiral arm is fitted by varying the pitch angle
(e.g., the Sagittarius arm in \citealt{1993ApJ...411..674T}).
In fact, \citet{2015ApJ...800...53H} recently showed that in
some external galaxies, the pitch angle in the same spiral arm varies,
and the variance is as large as that among different arms within the same galaxy.
So a more accurate assumption may be that the pitch angle of one segment of the arm within one galaxy is constant. 

Now we assume that the segment form $l=100^{\circ}$ to $l=150^{\circ}$ of 
the Outer arm spiral is a logarithmic spiral,
and using the same equation as \citet{2009ApJ...700..137R,2014ApJ...783..130R} 
we fitted the spiral of those 457 Outer arm clouds using
\begin{equation}\label{Eq: Pitch}
     \ln (\frac{R}{R_{\mathrm{ref}}})=-(\beta-\beta_{\mathrm{ref}})\tan(\psi)
\end{equation}
where $\psi$ is the spiral pitch angle, 
$\beta$ is the galactocentric azimuth (namely the Source-GC-Sun angle),
and $R_{\mathrm{ref}}$ and $\beta_{\mathrm{ref}}$ are the reference radius and reference azimuth, respectively.
The fitting algorithm is a minimizing chi-square error statistic,
and the chi-square error statistic is computed as
\begin{equation}
      \chi^2(k,b)=\sum_{i=1}^{N} W_i(y_i-b-kx_i)^2
\end{equation}
where $k=-\tan(\psi)$, $b= \beta_{\mathrm{ref}}\tan(\psi)+ln(R_{\mathrm{ref}})$,
$x_i$ and ${y_i}$ are $\beta$ and $\ln(R)$ of each cloud, respectively,
and $W_i$ is the weight.
The weight is defined as $W=\log_{10}(\frac{M}{M_{\odot}})$.

\begin{figure}[htbp]
\epsscale{0.7}
\plotone{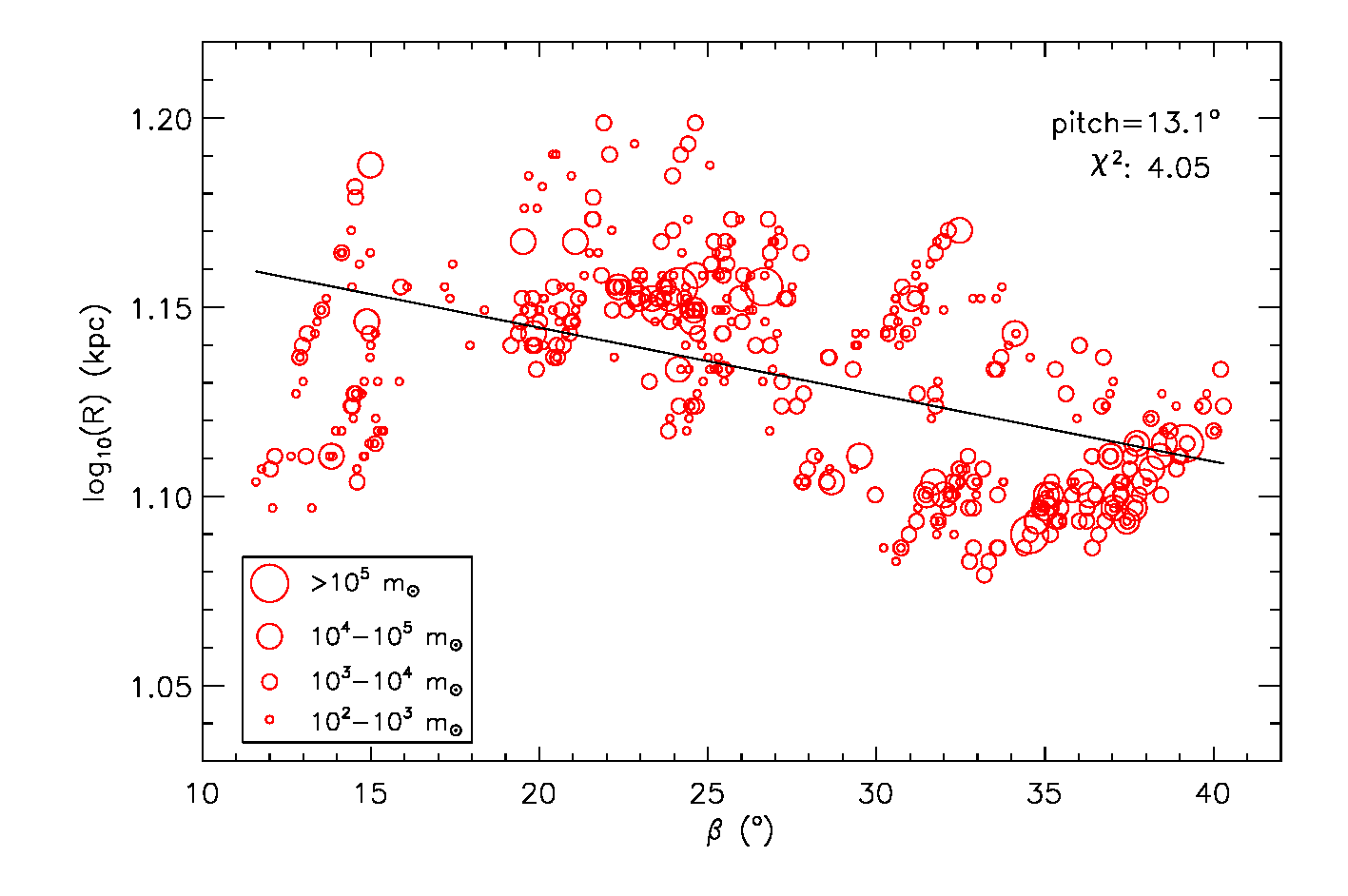}
\caption{Fitting result of the Outer arm pitch angle.
R is Galactocentric radius in the units of kpc.
$\beta$ is Galactocentric longitude.
Circles indicate molecular clouds of the Outer arm, and their sizes indicate different masses.
The fitting line is weighted by mass.
}\label{Fig: pitch fit}
\end{figure}

Fig.\ref{Fig: pitch fit} shows the fitting result. The pitch angle is $\sim 13.1^{\circ}$,
the $R_{\mathrm{ref}}$ is $\sim 13.6$ kpc,
the $\beta_{\mathrm{ref}}$ is $\sim 26.9^{\circ}$ and 
the $\chi^2$ is $\sim4.1$.
Our fitting pitch angle is close to the result of \citet{2014ApJ...783..130R} (pitch=$13.8^{\circ}$),
and is consistent with the results of \citet{2015MNRAS.450.4277V},
who summarized large numbers of recent studies about Milky Way pitch angle
and yielded a mean global value of $13.1^{\circ}$.

\clearpage
\subsection{Plan view}\label{SubSec: Plan view}
Since \citet{1976A&A....49...57G} presented the famous Milky Way plan view,
numerous models have been suggested
(see Figure 2 in \citealt{2010gama.conf...45S}).
Our knowledge about the Milky Way structure has advanced since then a little,
so at least the existence of Outer arm is now undoubted,
though it was absent in the Georgelin model.
In contrast, the lack of arm tracers, especially the remote tracers,
still prevent us from knowing more about our galaxy.
Thanks to the high-sensitivity CO survey of MWISP, so many molecular clouds of the Outer arm have been detected,
and this is the first time that we have detected such large numbers of remote molecular clouds,
which contributes a lot to studying the Milky Way structure.

\begin{figure}[htbp]
\epsscale{0.9}
\plotone{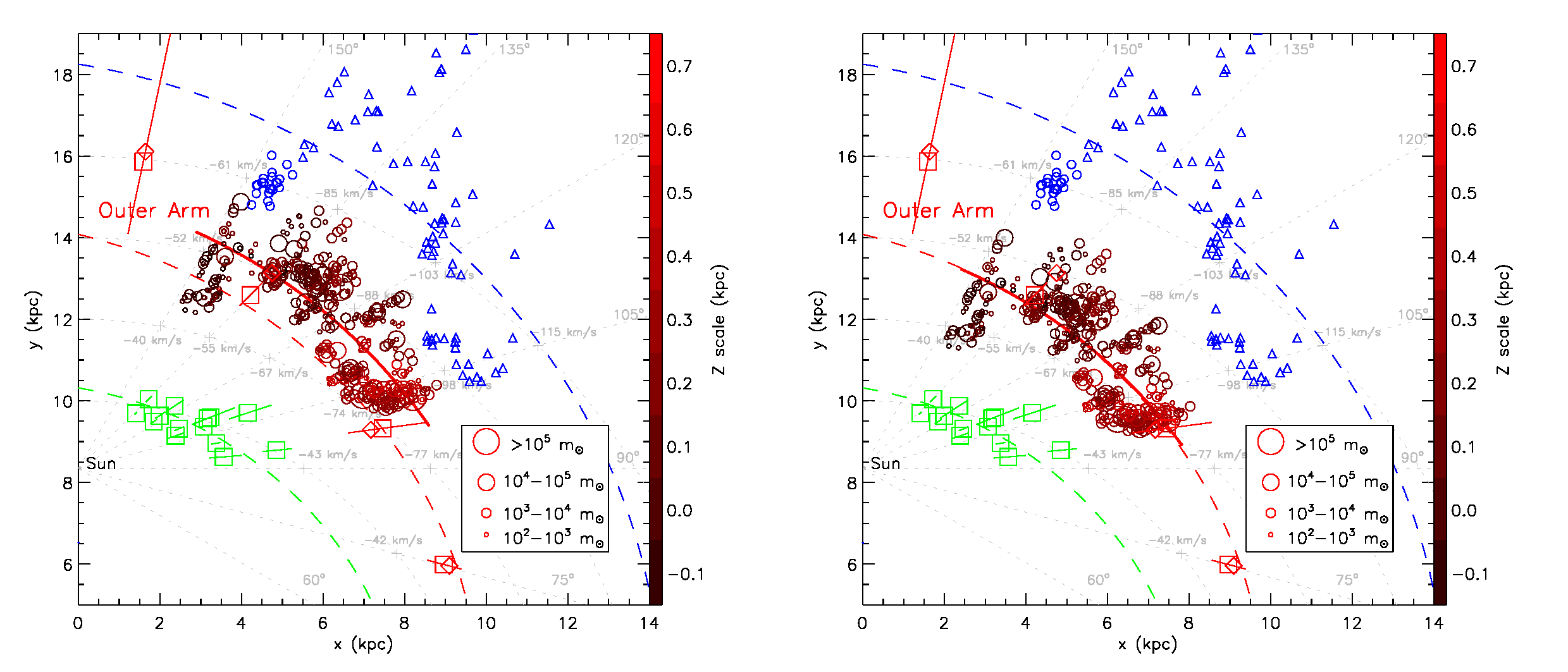}
\caption{(\emph{left panel}) Locations of molecular clouds and HMSFRs in the plan view of the Milky Way.
The red and blue circles respectively mark the molecular clouds of
the Outer arm and New arm, which are summarized in Table. \ref{Table}.
The blue triangles mark the molecular clouds of the New arm that were detected by \citet{2015ApJ...798L..27S}.
Different sizes of circles and triangles indicate different masses.
The red and green squares respectively mark the HMSFRs \citep{2014ApJ...783..130R} of Outer and Perseus arms.
Distance error bars of HMSFRs are indicated.
The color depths of the red circles indicate their $Z$ heights.
The red diamonds indicate the locations of the Outer arm HMSFRs calculated by kinematic method.
The green, red, and blue dashed curves respectively indicate 
the Reid Perseus spiral, the Reid Outer spiral, and the Sun New spiral.
The red thick curve indicates our Outer spiral.
The gray dotted curves indicate galactocentric radii=10, 12, 14, 16 kpc.
The gray dotted lines indicate galactic longitudes=60$^{\circ}$, 75$^{\circ}$, 90$^{\circ}$, 105$^{\circ}$, 120$^{\circ}$, 135$^{\circ}$, 150$^{\circ}$.
The gray crosses and the words beside indicate the locations and the corresponding LSR velocities.
(\emph{right panel}) Translational result of the Outer arm clouds and our Outer spiral.
For a better comparison, we have not moved or changed any symbols except the red circles and the red thick curve,
and the sizes of the red circles remain the same although their masses are changed because of the revised distances.
}\label{Fig: xymap-ab}
\end{figure}

The left panel of Fig. \ref{Fig: xymap-ab} shows the plan view of the Milky Way.
The red thick curve indicates the Outer arm spiral fitted by us (hereafter our Outer spiral). 
Since the Outer arm pitch angle fitted by us is similar to the one fitted by \citet{2014ApJ...783..130R},
and their parallax distance of the Outer arm is more precise,
we moved our Outer spiral to be parallel to the position of the Reid Outer spiral.
Accordingly, the Outer arm clouds are parallelly moved by the same distance.
The detailed moving process is as follows:
First, using the $R_{\mathrm{ref}}$ and $\beta_{\mathrm{ref}}$ fitted by \citet{2014ApJ...783..130R}
with the pitch angle fitted by us, we plotted a new spiral curve as the parallel moved curve.
Second, we calculated the distance between the new curve and the old curve at every cloud galactic longitude.
Third, the new cloud heliocentric distance was calculated as the
``old cloud heliocentric distance minus the distance obtained in the second step''.
This moving process decreased the cloud heliocentric distances but did not change their longitudes.
The right panel of Fig. \ref{Fig: xymap-ab} is the result after the move.

We mentioned these distance biases in \ref{SubSec: Cloud parameters}
One may wonder why the kinematic distances
obtained from the model that \citet{2014ApJ...783..130R} provided are greater than
the distances that they measured.
In fact, the biases are mainly caused by errors.
Limited parallax data and the parallax measuring errors lead to larger errors of the galactic model.
The Reid model narrows the gap between the kinematic distance and the parallax distance but cannot perfectly match them.
(And actually, the cloud distances that we moved are roughly below the parallax distance errors.
Just see the error bars of the Outer arm HMSFRs in Fig. \ref{Fig: xymap-ab}.)
Besides, the Reid model is a global model and
it is fitted by the maser sources that distribute in almost three galactic quadrants
(see Figure 1 of \citet{2014ApJ...783..130R}).
But different regions of the Milky Way may have their own peculiar motions.
So in different regions the kinematic distances calculated from their model may be a little biased.
As a result of these reasons,
in the second galactic quadrant (or more exactly in the Outer arm region of the second galactic quadrant)
the kinematic distances are a little larger than the parallax distances,
just as we see in the left panel of Fig. \ref{Fig: xymap-ab}.

\begin{figure}[htbp]
\epsscale{0.7}
\plotone{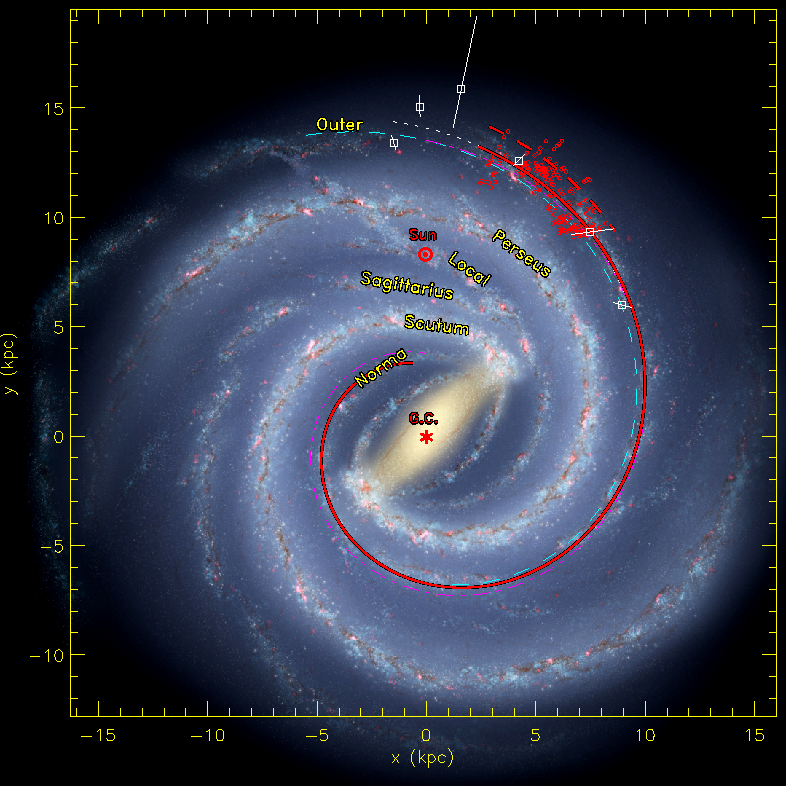}
\caption{An artist's conception of the Milky Way (R. Hurt: NASA/JPL-Caltech/SSC).
The red circles indicate the parallel moved Outer arm clouds with mass $\rm > 10^{3}M_{\odot}$.
The white squares indicate the Outer arm HMSFRs and the distance error bars are also plotted.
The red thick curve indicates the extended translational position of our Outer spiral,
and the red thick dashed curve indicates the original position of our Outer spiral.
The white dotted curve indicates the Reid Outer spiral.
The cyan dashed curve and magenta pecked curve indicate the
Outer arm fitting results of \citet{2014A&A...569A.125H} and \citet{2014MNRAS.437.1549B}, respectively.
}\label{Fig: xymap-art}
\end{figure}

Fig. \ref{Fig: xymap-art} shows the whole view of the Milky Way.
We extended our translational (namely the parallel moved) spiral to the inner galaxy.
The Reid Outer spiral is also plotted.
Additionally, for comparison we plotted two recent fitting results for this arm,
of which the arm tracers or fitting methods are different.
The cyan dashed curve is one of the fitting results from \citet{2014A&A...569A.125H}
(``arm-5'' in the third column of Table 4 in their paper).
Their arm tracers are HII regions, and their fitting model is the polynomial-logarithmic spiral arm model.
The magenta pecked curve is the fitting result from \citet{2014MNRAS.437.1549B}.
Their arm tracers are 3 HMSFRs and 12 very young star clusters,
and their fitting model is the the logarithmic spiral.
The fact that our translational spiral is located closer to the other three spirals may suggest that
the translational locations of the clouds are relatively better.
However, this suggestion is not adequate enough to
make us revise the distances and other parameters such as mass in Table. \ref{Table}.
The original parameters derived in Sect. \ref{SubSec: Cloud parameters} are retained.

\clearpage
\subsection{\textit{l-V}, \textit{V-b} and \textit{l-b} map}\label{SubSec: l-v, v-b, l-b map}
\begin{figure}[htbp]
\epsscale{1.0}
\plotone{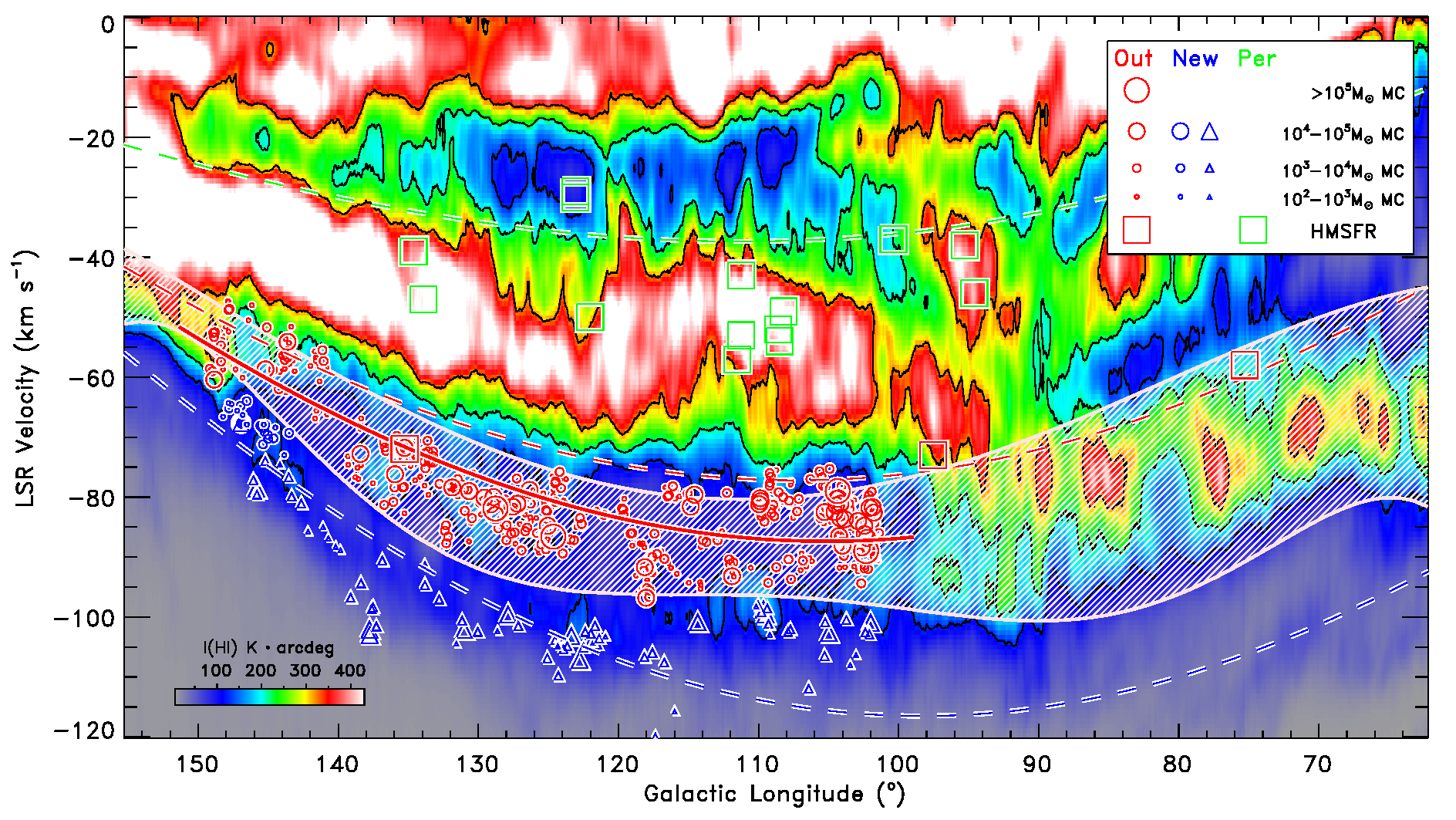}
\caption{Longitude-velocity diagram of HI from the CGPS, integrated over all latitudes 
(from $-3^{\circ}$ to $5^{\circ}$).
The red and blue circles respectively mark the molecular clouds of 
the Outer arm and the New arm, which are summarized in Table. \ref{Table}.
The blue triangles mark the molecular clouds of the New arm detected by \citet{2015ApJ...798L..27S}.
Different sizes of circles and triangles indicate different masses.
The red and green squares respectively mark the HMSFRs of the Outer and Perseus arms.
The green, red, and blue dashed curves respectively indicate the projections of 
the Reid Perseus spiral, the Reid Outer spiral, and the Sun New spiral.
The red thick curve indicates the projection of the Outer arm spiral fitted by us.
The pink ribbon indicates the integrated velocity window of Fig. \ref{Fig: l-b map}.
}\label{Fig: l-v map}
\end{figure}

Fig. \ref{Fig: l-v map} shows the \textit{l-V} map of HI emission.  
All the clouds (including the New arm clouds) and HMSFRs are marked on it.
Also, the projections of the Reid Outer spiral, the Reid Perseus spiral, the Sun New spiral, and
our Outer spiral are plotted.
In order to project those spiral curves on the \textit{l-V} map,
or in other words, to convert the dashed curves shown in Fig. \ref{Fig: xymap-ab}
into the ones shown in Fig. \ref{Fig: l-v map},
we compiled a C program on the basis of the FORTRAN code provided by \citet{2009ApJ...700..137R}.
Their program can calculate the revised heliocentric kinematic distances
when inputting the cloud positions and the LSR velocities,
just as mentioned in Sect. \ref{SubSec: Cloud parameters}.
Our program is the inverse --- the inputs are cloud position
(galactic longitude and latitude) and galactocentric radius, and the output is the LSR velocity.
Knowing the expression of the arm curves (namely through Eq. \ref{Eq: Pitch}),
we can obtain an array of galactocentric radii at every galactic longitude.
And then ,using the Reid model, the cooresponding LSR velocities can be calculated by our C program.
Using the arrays of longitudes and velocities, we then can plot \textit{l-V} curves of the arms.

\begin{figure}[htbp]
\epsscale{1.0}
\plotone{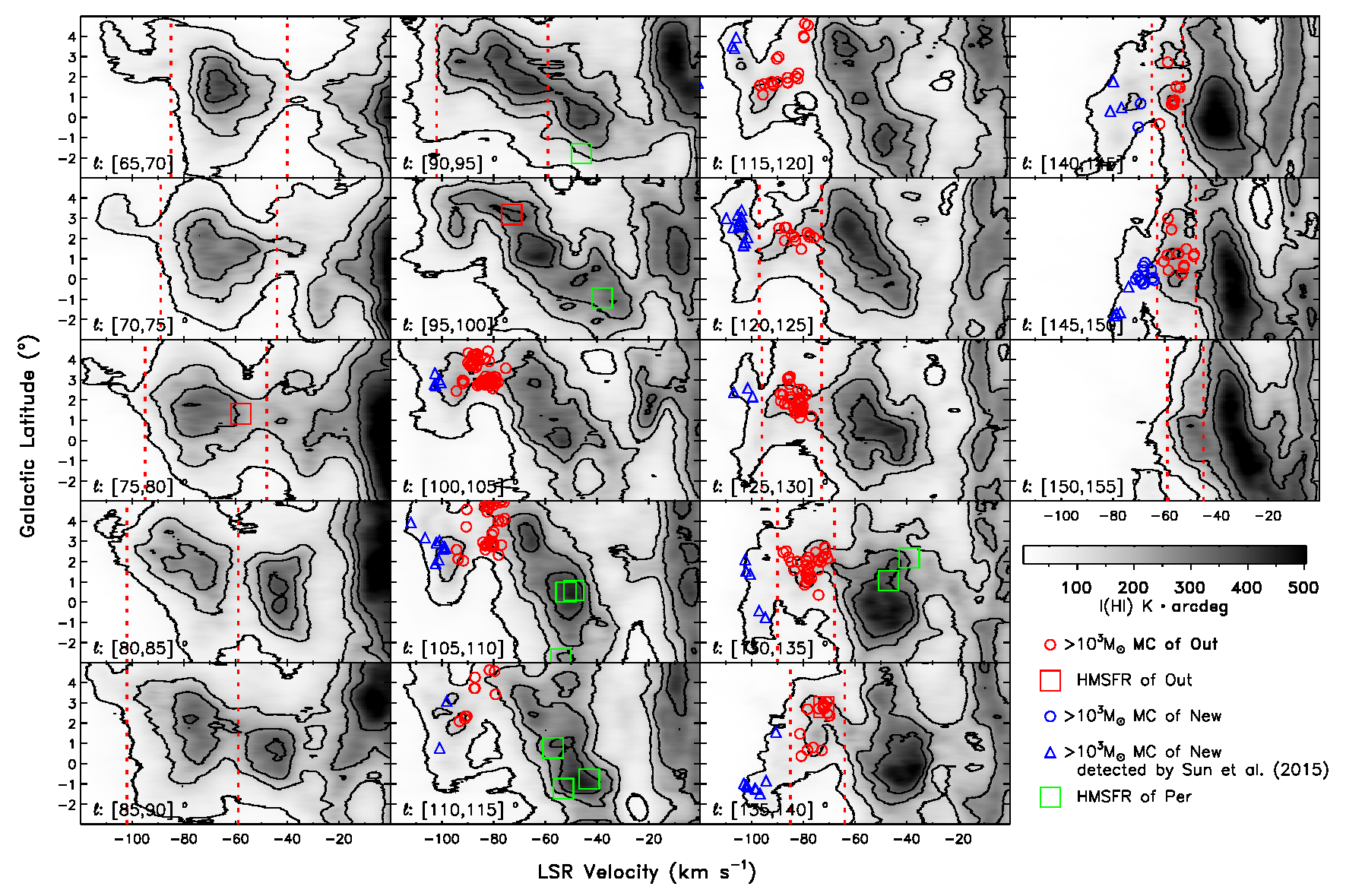}
\caption{Velocity-latitude montage of HI from the CGPS, integrated every $5^{\circ}\!$ of galactic longitude.
The Outer arm is embraced in the two dotted red lines in each figure
except the figures of the arm-blending region.
The meanings of other symbols are consistent with Fig. \ref{Fig: l-v map}.
Note that the low-mass clouds($\rm < 10^{3}M_{\odot}$) are not plotted, 
and the symbol size does NOT indicate mass.
}\label{Fig: b-v map}
\end{figure}

Fig. \ref{Fig: b-v map} shows the \textit{V-b} montage of HI emission.
The high-mass Outer arm clouds ($\rm >10^{3} M_{\odot}$),
the New arm clouds, and the HMSFRs are marked on it.

As mentioned in Sect. \ref{SubSec: Cloud identification},
from longitude $l\simeq100^{\circ}$ to $120^{\circ}$, there exists an arm-blending region.
This phenomenon can be clearly seen from those two figures
and may mainly be caused for the following three reasons: 
(i) Streaming motions near spiral arms have long been predicted by density wave theory
and has been observed in other galaxies.
 (e.g. Figure 4 in \citealt{1980A&A....88..159V}, Figure 5 in \citealt{1999ApJ...522..165A})
This can probably happen at this region in our galaxy.
(ii) Spiral shock may lead to the condition that
one LSR velocity could share two different distances.
(e.g., \citealt{2006ApJ...644..214F})
(iii) The expansion motion of the HI super-bubble near $l=123^{\circ}$, $b=-6^{\circ}$
associated with the Perseus arm may also lead to the velocity mixing. \citep{2007PASJ...59..743S}
Whatever reason mainly causes the mixing LSR velocities,
this at least suggests that in this region the gas motion is very peculiar.

\begin{figure}[htbp]
\epsscale{1.0}
\plotone{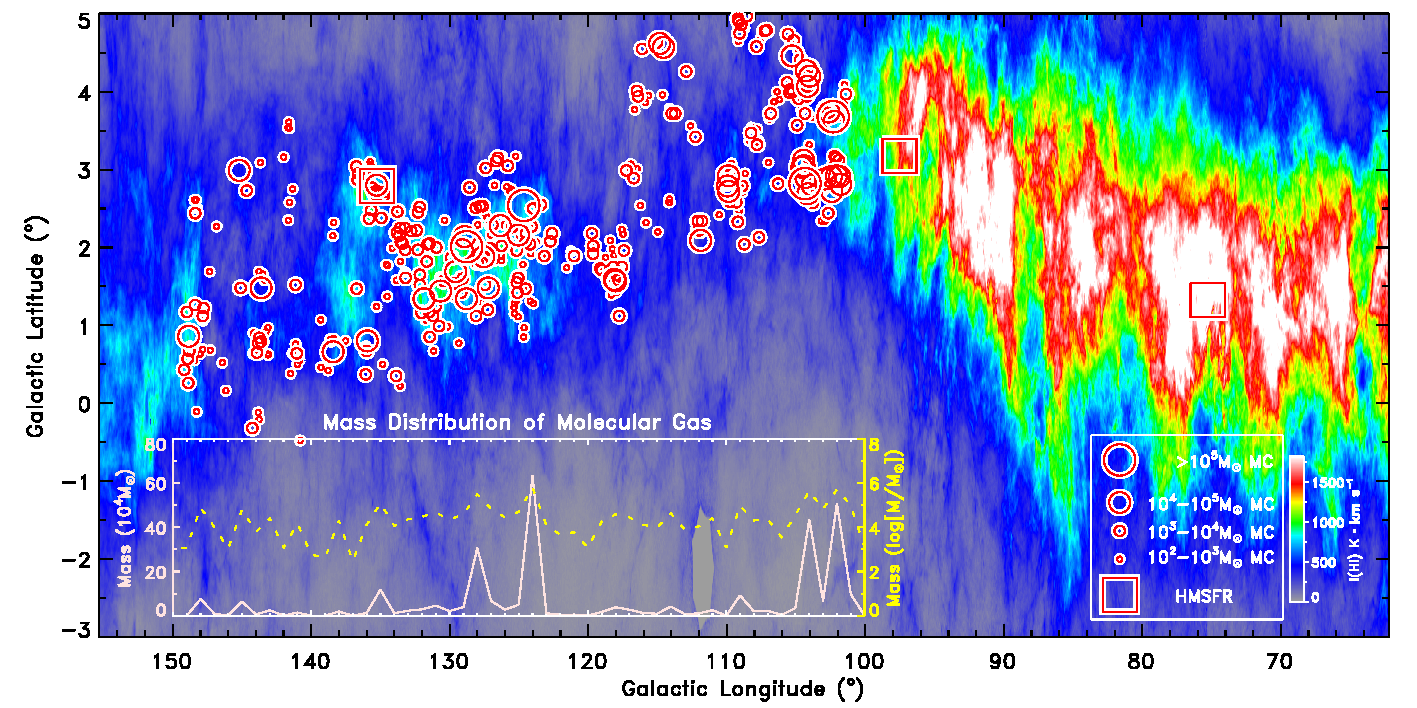}
\caption{Velocity integrated intensity of HI from the CGPS corresponding to the Outer arm. 
The integrated velocity window is marked by the pink ribbon in Fig. \ref{Fig: l-v map}.
The meanings of the symbols are consistent with Fig. \ref{Fig: l-v map}.
The inset shows the mass of molecular gas as a function of galactic longitude: 
the solid line and the dashed line correspond to different units.
}\label{Fig: l-b map}
\end{figure}

Fig. \ref{Fig: l-b map} shows the velocity-integrated intensity of HI emission.
All the clouds and HMSFRs of the Outer arm are marked on it.
Since the velocity ranges of the Outer arm are different at different longitudes,
we need to define an $l-V$ function as the integrated velocity window.
We have adopted two polynomial fitting curves as the outline of integrated range instead of the spiral $l-V$ projection.
This is because the Outer arm velocity ranges are irregular
and the polynomial fitting curve is more appropriate for defining the various ranges.
The inset of Fig. \ref{Fig: l-b map} shows the mass of molecular gas as a function of galactic longitude.
The bin is 1 galactic longitude degree.
The total mass of all the Outer arm clouds is $\rm \sim 3.1\times10^{6} \ M_{\odot}$.
More details about gas distributions are presented in Sect. \ref{SubSec: Gas distribution}.

Those three figures present a 3D view of the arm.
The HI and H$_{2}$ gases are roughly matched.
The warp is obvious.

\clearpage
\subsection{Gas distribution}\label{SubSec: Gas distribution}
The gas distribution of the Milky Way is a frequently discussed topic.
(e.g., \citealt{1976ApJ...208..346G}; \citealt{1987ApJ...322..101S};
\citealt{2003PASJ...55..191N} and their serial papers;
\citealt{2015MNRAS.447.2144D}).
Whereas, most works often discuss its global distribution (e.g., \citealt{1975ApJ...202...30B}),
only a few have focused on the individual arm (e.g., \citealt{1987ApJ...315..122G}). 
Now we present our results of the Outer arm gas distribution.

\begin{figure}[htbp]
\epsscale{0.7}
\plotone{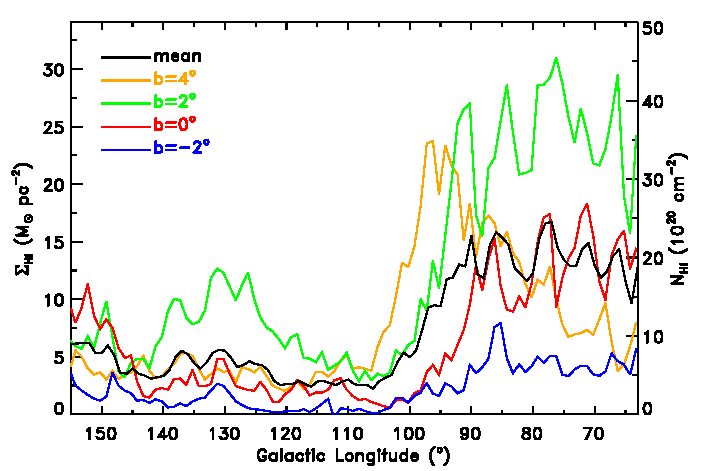}
\caption{Distribution of HI surface density (or column density) of the Outer arm along the galactic longitude.
The lines with different colors indicate different galactic latitudes,
and the black line indicates the mean value.
}\label{Fig: HI surface density}
\end{figure}

Fig. \ref{Fig: HI surface density} shows the Outer arm HI surface density distribution
(or column density distribution) along the galactic longitude.
Assuming that the 21 cm line is optically thin, the HI surface density is calculated from
$\Sigma_{\mathrm{HI}}=1.82\times10^{18}m_{\mathrm{H}}\int T_{\mathrm{B}}d \,V$,
and the column density is calculated by the conversion factor of
$\mathrm{1 \ M_{\odot}pc^{-2}=1.25\times10^{20} \ \ cm^{-2}}$,
where $ m_{\mathrm{H}}$ is the H atom mass,
and $\int T_{\mathrm{B}} d \,V$ is the integrated intensity.
The velocity-integrated range is the pink ribbon shown in Fig. \ref{Fig: l-v map}.
Clearly, the HI surface density abruptly descends at $l\simeq100^{\circ}$,
and ascends a little at $l\simeq120^{\circ}$.
The arm-blending region may account for the sharp drop in $l=[100,120]^{\circ}$,
but cannot explain the low surface density region of $l=[120,155]^{\circ}$.
This may suggest that in one spiral arm the gas quantity can largely change.
One possible reason may be that:
some of the apparent variation in HI surface density could arise from colder, optically thick HI gas,
which is known to be widespread in the outer Galaxy,
perhaps overlapping with some of the molecular clouds.
(see \citealt{2001Natur.412..308K}, \citealt{2007AJ....134.2252S}
and \citealt{2010ASPC..438..111G})

\begin{figure}[htbp]
\epsscale{0.7}
\plotone{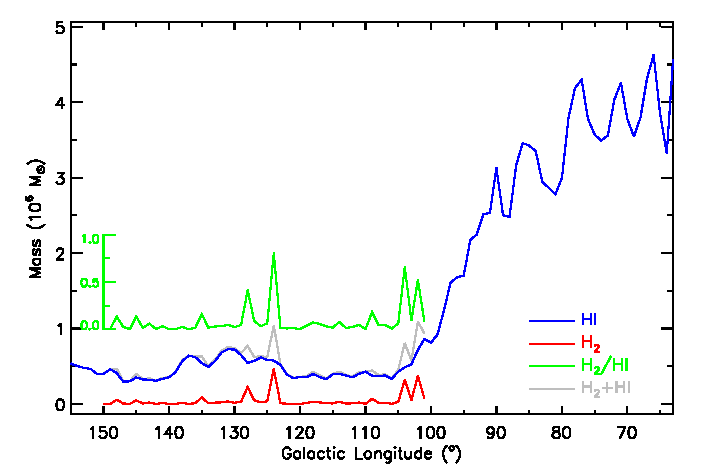}
\caption{Distribution of HI \& H$_{2}$ masses of the Outer arm along the galactic longitude.
}\label{Fig: mass distribution}
\end{figure}

\begin{figure}[htbp]
\epsscale{0.7}
\plotone{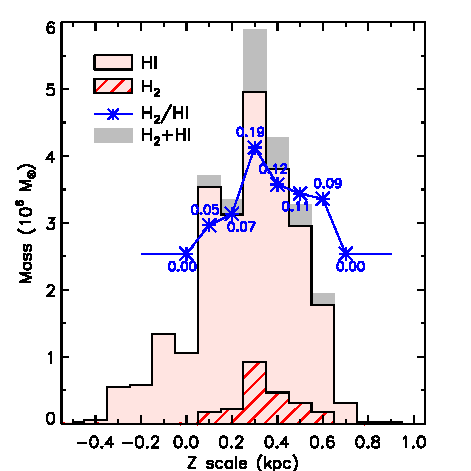}
\caption{Distribution of the HI \& H$_{2}$ masses along the scale height.
This result only shows the region $l=[100,150]^{\circ}$, $b=[-3,5]^{\circ}$.
The blue words beside the asterisks are the mass ratios of H$_{2}$ to HI.
}\label{Fig: gas distribution in Z}
\end{figure}

Using the HI surface density and kinematic distance, the HI mass is obtained
from the following process.
First, since we know the HI surface density at every pixel of the HI integrated map
(namely Fig. \ref{Fig: l-b map}),
we then can calculate the mean surface density at every square degree.
(This step is somewhat like smoothing the integrated map into a one square degree per pixel map.)
Second, using the Reid model we calculate the kinematic distance of every square degree,
and the mean LSR velocity of the integrated window (namely the pink ribbon of Fig. \ref{Fig: l-v map})
at every galactic longitude is used as the LSR velocity of its corresponding square degree.
Third, using the kinematic distance and the surface density
we can calculate the HI mass in every square degree by
$M_{\mathrm{HI}}=\Sigma_{\mathrm{HI}}d^{2}_{\mathrm{HI}}$,
where $\Sigma_{\mathrm{HI}}$ and $d_{\mathrm{HI}}$
respectively indicate the surface density and kinematic distance at the corresponding square degree.
Fig. \ref{Fig: mass distribution} shows the HI and H$_{2}$ mass distribution along the galactic longitude.
(Note that H$_{2}$ is different from molecular gas; the latter includes helium.)
The bins of HI and H$_{2}$ are both 1 galactic longitude degree.
The total mass of HI in $l=[63,155]^{\circ}$, $b=[-3,5]^{\circ}$ of the Outer arm is 
$\rm\sim 1.4\times10^{8} \ M_{\odot}$.
And the total mass of H$_{2}$ in $l=[100,150]^{\circ}$, $b=[-3,5]^{\circ}$ is 
$\rm \sim 2.3\times10^{6} \ M_{\odot}$.
Since the observation is not fully covered in $b$, the mass must be higher.
The mean mass ratio of H$_{2}$ to HI in $l=[100,150]^{\circ}$, $b=[-3,5]^{\circ}$ is about 0.1.
Fig. \ref{Fig: gas distribution in Z} shows the HI and H$_{2}$ mass distributions along the $Z$ scale 
in the region of $l=[100,150]^{\circ}$, $b=[-3,5]^{\circ}$.
About 50\% of the gas mass is included in Z=[0.2,0.4] kpc.

Interestingly, there seems to be a trend in the $Z$ scale distribution (Fig. \ref{Fig: gas distribution in Z}):
as the total gas mass becomes large, the H$_{2}$ to HI ratio becomes higher.
For example, in Fig. \ref{Fig: gas distribution in Z},
at $Z=0.3$ kpc, the total gas mass is the largest, and accordingly the H$_{2}$ to HI ratio is the highest.
On the other hand, at $Z=0.1$ or 0.6 kpc, the gas mass descends and so does the ratio.
Also, this trend is visible in the longitude distribution (Fig. \ref{Fig: mass distribution}).
One possible explanation may be that
although the H$_{2}$ distribution is not as diffuse as HI,
once it exists, it contributes a lot to the total mass.

It is necessary to emphasize that
we are reporting only the mass of H$_{2}$ traced by detected CO emission.
Significant additional H$_{2}$ which CO cannot detect
(or say the ``CO-dark'' H$_{2}$, e.g., \citealt{2005Sci...307.1292G}) may also be present.
Most of the sight lines plotted in Fig. \ref{Fig: HI surface density}
exceed the minimum column density for self-shielding H$_{2}$
(a few times 10$^{20}$ cm$^{-2}$;
see \citealt{2006ARA&A..44..367S} and \citealt{2008ApJ...687.1075S}).
This is for individual clouds rather than integrated sight lines,
but it still seems likely that H$_{2}$ could exist in many sightlines where the total HI column is high enough.

\clearpage
\subsection{Warp and thickness}\label{SubSec: Warp and thickness}
Warp is a general phenomenon among disk galaxies.
(e.g., \citealt{1976A&A....53..159S}; \citealt{1990MNRAS.246..458S};
\citealt{1998A&A...337....9R}; \citealt{2002A&A...394..769G}).
Additionally, it is one of the most fascinating and important subjects in the study of the Milky Way structure
(e.g., \citealt{1957AJ.....62...90B}; \citealt{1996AJ....111..804R};
\citealt{2006A&A...451..515M}; \citealt{2010HiA....15..811L}).
Figure 3 in the review of \citet{2009ARA&A..47...27K} shows a distinct three-dimensional (3D) picture of the warped Milky Way plane.

\begin{figure}[htbp]
\epsscale{0.7}
\plotone{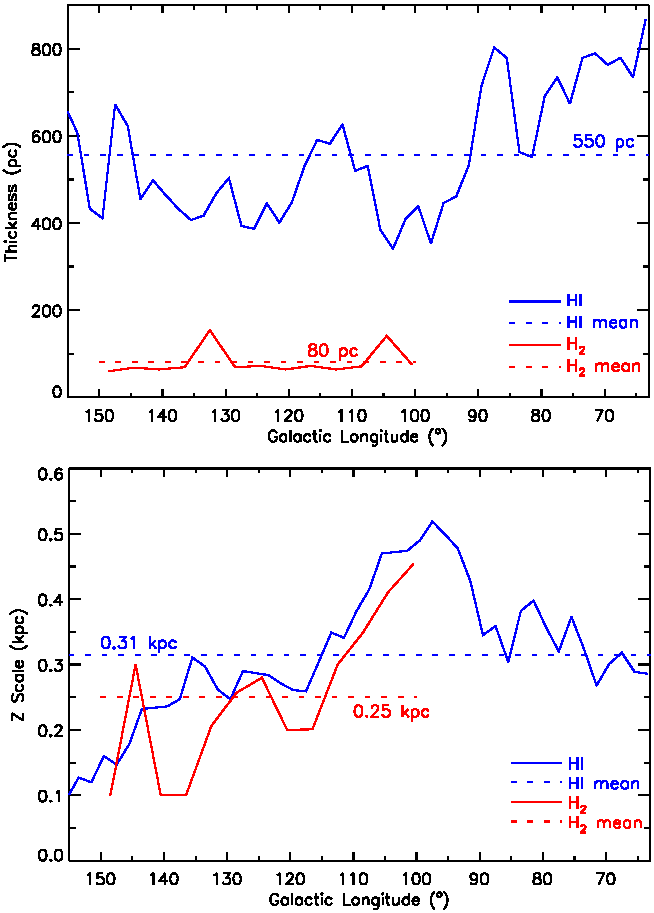}
\caption{Distribution of the Outer arm thickness (\emph{Upper panel}) 
and scale height (\emph{lower panel}) along the galactic longitude.
}\label{Fig: thick and z scale distribution}
\end{figure}

In our paper, one can easily notice the Milk Way warp from
Fig. \ref{Fig: b-v map}, \ref{Fig: l-b map}, \ref{Fig: HI surface density} and \ref{Fig: gas distribution in Z}.
We can roughly obtain a first impression from Fig. \ref{Fig: l-b map}:
the Outer arm starts to bend upward at $l\simeq80^{\circ}$,
and at $l\simeq105^{\circ}$ it reaches its peak and then begins to fall.
The whole arm from $l=63^{\circ}$ to $l=155^{\circ}$ just looks like an arch.
In addition, almost all segments of this arch are beyond the $b=0^{\circ}$ plane.
But since the distance is different at every longitude,
the \textit{l-b} map cannot fully represent the scale height distribution.
Now a more detailed analysis is presented.
First, we plotted HI mass distributions along the $Z$ scale every 2 galactic longitudes,
and H$_{2}$ mass distributions every 4 galactic longitudes.
Then we fitted every distribution with a Gaussian curve.
We define the centric position as the $Z$ scale and the FWHM is the thickness.
The final results are shown in Fig. \ref{Fig: thick and z scale distribution}.
The mean $Z$ scales of the HI layer and the H$_{2}$ layer are about 0.31 kpc and 0.25 kpc, respectively.
The mean thicknesses of the HI layer and the H$_{2}$ layer are about 550 pc and 80 pc, respectively.
The $Z$ scales of HI and H$_{2}$ are close ($\simeq$ 0.3 kpc),
but the HI thickness is much larger than that of H$_{2}$
(about 7 times thicker, which is not obvious in Fig. \ref{Fig: l-b map}).

The increasing trend of spiral arm thickness with galactocentric radius has been widely observed and accepted.
(e.g., \citealt{1990A&A...230...21W}; \citealt{2009ARA&A..47...27K}).
Here we detected a similar trend.
Assuming a mean distance of 2 kpc,the HI thickness of the Perseus arm is about 200 pc corresponding to $5^{\circ}$.
And as it mentioned above, the Outer arm thickness is about 550 pc.
Meanwhile, the thickness of the New arm is 400 to 600 pc \citep{2015ApJ...798L..27S}.
Obviously the Outer arm is much thicker than the Perseus arm, but is nearly as thick as the New arm.
Maybe this provides a little evidence to a trend newly discovered by \citet{2015ApJ...800...53H}
that in the outermost parts of the galaxies some arms become narrow.

\clearpage
\section{Summary}\label{Sec: Summary}
Combining 115 GHz $^{12}$CO(1-0) data ($l=[100,150]^{\circ}$, $b=[-3,5]^{\circ}$ ) of MWISP
and 21 cm HI data ($l=[65,115]^{\circ}$, $b=[-3,5]^{\circ}$) of CGPS,
we present the properties of Outer arm in the second galactic quadrant of the Milky Way.

(1) Using CO(1-0), We have detected 481 molecular clouds in total,
of which 332 (about 69\%) are newly detected,
457 clouds are identified in the Outer arm, and 24 are identified in the New arm.
The parameters of all the 481 clouds are summarized in Table. \ref{Table}.

(2) Assuming that the spiral arm is logarithmic spiral,
the pitch angle of the Outer arm is fitted by minimizing chi-square error statistic algorithm,
and the result is $\sim$ $13.1^{\circ}$.

(3) The total masses of molecular gas and H$_{2}$ from $l=100^{\circ}$ to $150^{\circ}$
in the Outer arm are about $\rm 3.1\times10^{6} \ M_{\odot}$ 
and $\rm 2.3\times10^{6} \ M_{\odot}$respectively.
And the total mass of HI gas from $l=63^{\circ}$ to $155^{\circ}$ is about $\rm 1.4\times10^{8} \ M_{\odot}$
Since the observation is not fully covered in $b$, the mass must be higher.
The mean mass ratio of H$_{2}$ to HI in $l=[100,150]^{\circ}$, $b=[-3,5]^{\circ}$ is about 0.1.

(4) The warp of the Outer arm is obvious.
The mean Outer arm thicknesses of HI and H$_{2}$ are about 550 and 80 pc, respectively,
while the scale height of both gases is $\simeq$ 0.3 kpc.

\acknowledgments
We are grateful to all the members of the Milky Way Scroll Painting survey group, 
especially the staff of the Qinghai Radio Observing Station at Delingha for technical support.   
Additionally, we gratefully acknowledge the anonymous referee for the helpful and valuable comments.
This work is supported by the National Natural Science Foundation of China
(grant numbers: 11133008 and 11233007), the Strategic Priority Research Program
of the Chinese Academy of Sciences (grant number: XDB09010300),
and the Key Laboratory for Radio Astronomy.

\clearpage
\begin{deluxetable}{lccccccccccl}
\tabletypesize{\scriptsize}
\tablewidth{0pt}
\tabcolsep = 1pt
\tablecaption{\label{Table}Parameters of molecular clouds derived by CO(1-0)}
\tablehead{
\colhead{Name}                                              & 
\colhead{$V_{\mathrm{LSR}}$}  &\colhead{$T_{\mathrm{peak}}$}          & 
\colhead{$\Delta V$} &\colhead{$I_{\mathrm{CO}}$}            &
\colhead{Area}           &\colhead{$d$}                       &
\colhead{$R$}              &\colhead{Size}                    &
\colhead{Mass}           &\colhead{$Z$ Scale}                 &
\colhead{Note}                                             \\

\colhead{}                                                  &
\colhead{(km $\rm s^{-1}$)}                                 & 
\colhead{(K)}            &\colhead{(km $\rm s^{-1}$)}       & 
\colhead{($\rm K \cdot$ km $\rm s^{-1}$)}                   &
\colhead{($\rm arcmin^{2}$)}                                &
\colhead{(kpc)}          &\colhead{(kpc)}                   &
\colhead{(pc)}           &\colhead{($\rm 10^{3}M_{\odot}$)} &
\colhead{(kpc)}          &\colhead{}                       \\ 

\colhead{(1)}            &\colhead{(2)}                     &
\colhead{(3)}            &\colhead{(4)}                     & 
\colhead{(5)}            &\colhead{(6)}                     & 
\colhead{(7)}            &\colhead{(8)}                     &
\colhead{(9)}            &\colhead{(10)}                    &
\colhead{(11)}           &\colhead{(12)}              
}
\startdata
MWISP G101.400+03.967  &  -86.0  &  2.3  &  1.6  &  3.9  &  15.4  &  8.2  &  12.8  &  10.4  &  1.3  &  0.6  &  \\    
MWISP G101.508+04.092  &  -90.5  &  3.2  &  1.1  &  3.6  &  3.0  &  8.6  &  13.1  &  4.4  &  0.2  &  0.6  &  \\    
MWISP G101.633+02.867  &  -82.6  &  3.6  &  1.6  &  6.2  &  10.0  &  7.8  &  12.5  &  7.9  &  1.2  &  0.4  &  \\    
MWISP G101.683+02.917  &  -81.7  &  4.4  &  1.3  &  5.9  &  21.7  &  7.7  &  12.4  &  11.6  &  2.5  &  0.4  &  \\    
MWISP G101.700+03.833  &  -89.9  &  7.9  &  1.4  &  11.5  &  25.8  &  8.6  &  13.1  &  14.2  &  7.4  &  0.6  &  \\    
MWISP G101.708+02.867  &  -84.3  &  4.9  &  1.2  &  6.3  &  18.4  &  8.0  &  12.6  &  11.1  &  2.5  &  0.4  &  \\    
MWISP G101.767+02.808  &  -82.6  &  16.5  &  2.9  &  51.0  &  43.3  &  7.8  &  12.5  &  16.7  &  45.8  &  0.4  &  \\    
MWISP G101.800+02.958  &  -81.4  &  2.5  &  0.7  &  2.0  &  4.8  &  7.7  &  12.4  &  5.2  &  0.2  &  0.4  &  \\
\enddata
\tablecomments{
Column (1): source named by MWISP project and $\rm T_{peak}$ position in Galactic Coordinate.
Columns (2)$-$(5): results of Gaussian fit to the spectra.
Column (6): solid angle defined by the 3$ \sigma$ limits. 
Columns (7)$-$(8): heliocentric distance and galactocentric radius; 
both are derived by galactic parameters obtained by \citet{2014ApJ...783..130R}. 
Column (9): equivalent diameters of the molecular clouds corrected by the beam size of the telescope.
Column (10): cloud mass calculated by X=$\rm 1.8 \times 10^{20} cm^{-2} (K \cdot km \ s^{-1})^{-1}$ \citep{2001ApJ...547..792D}.
Column (11): scale height (=Dsin(b)).
Column (12): BW94: sources detected by \citet{1994A&AS..103..503B}; HCS01: sources detected by \citet{2001ApJ...551..852H};
BKP03: sources detected by \citet{2003ApJS..144...47B}; 
New Arm: clouds supposed to belong to the new spiral arm detected by \citet{2015ApJ...798L..27S}.
}
\end{deluxetable}

\end{document}